\documentclass{aa}
\usepackage{txfonts}

\usepackage{psfig,graphics,astrobib,amssymb}

\newcommand{\be}{\begin{equation}}
\newcommand{\e}{\end{equation}}
\newcommand{\bear}{\begin{eqnarray}}
\newcommand{\ear}{\end{eqnarray}}

\newcommand{\f}{\frac}
\newcommand{\de}{{\rm d}}

\begin{document}

\title
{Cosmological parameters from supernova observations: A critical comparison 
of three data sets}
\author
{T. Roy Choudhury \inst{1}
\and
T. Padmanabhan \inst{2}}
\offprints{T. Roy Choudhury, \email{chou@sissa.it}}
\institute{SISSA/ISAS, via Beirut 2-4, 34014 Trieste, Italy\\
\email{chou@sissa.it}
\and
IUCAA, Ganeshkhind, Pune, India 411 007\\
\email{nabhan@iucaa.ernet.in}
}

\date{\today}

\abstract{
We extend our previous analysis of cosmological supernova Type Ia data 
\cite{pc03} to include three recent compilation of data sets. 
Our analysis ignores the possible correlations and systematic effects 
present in the data and concentrates mostly on some key theoretical 
issues.
Among the three data sets, the first set 
consists of 194 points obtained from various 
observations while the second 
discards some of the points from the 
first one because of large uncertainties 
and 
thus consists of 142 points. The third data set is obtained 
from the second by adding the latest 14 points observed through HST.
A careful comparison of these different data sets help us to draw 
the following conclusions: (i) 
All the three data sets strongly rule out 
non-accelerating models. 
Interestingly, the first and the second data sets favour a closed
universe; if $\Omega_{\rm tot}\equiv \Omega_m+\Omega_{\Lambda}$, then the
probability of obtaining models with $\Omega_{\rm tot} > 1$ is 
$\gtrsim 0.97$. 
Hence these data sets are in
mild disagreement with the ``concordance'' flat model. However, this
disagreement is reduced (the probability of obtaining models with
$\Omega_{\rm tot} > 1$ being $\approx 0.9$) for the third data set, which
includes the most recent points observed by HST around $1 < z < 1.6$.
(ii) When the first data set is divided into two separate 
subsets consisting of low ($z < 0.34$) and high ($z > 0.34$) 
redshift supernova, it 
turns out that these two subsets, individually, admit 
non-accelerating models with 
zero dark energy because of different magnitude zero-point values 
for the different subsets. 
This can also be seen when the data is analysed while allowing 
for possibly different
magnitude zero-points for the two redshift subsets.
However, the non-accelerating models 
seem to be ruled out using {\it only} the low redshift data for 
the other two data sets, which have 
less uncertainties. 
(iii) We have also 
found that it is quite difficult to measure the 
evolution of the dark energy equation of state $w_X(z)$ though 
its present value can be constrained quite well. 
The best-fit value seems to {\it mildly} favour a dark energy component
with current equation of state $w_X < -1$, thus opening 
the possibility of existence of more exotic forms of matter. However, 
the data is  still consistent with the
the standard cosmological constant at 99 per cent confidence level
for $\Omega_m \gtrsim 0.2$.
\keywords{supernovae: general -- cosmology: miscellaneous --
cosmological parameters}
}

\titlerunning{Cosmological parameters from supernova observations}
\authorrunning{Choudhury \& Padmanabhan}

\maketitle

\section{Introduction}

Current cosmological observations, particularly those 
of supernova Type Ia, show a strong signature of 
the existence of a 
dark energy component with negative pressure 
\cite{rfc++98,pag++99,riess00}.  The most obvious
candidate for this dark energy is the cosmological constant (with 
the equation of state $w_X = p/\rho = -1$), which, however, 
raises several theoretical difficulties [for reviews, 
see \citeN{ss00}, \citeN{pr03}, \citeN{padmanabhan03}].
This has led to 
models for  dark energy component 
which evolves with time \cite{rp88,wetterich88,fj98,fhsw95,bm99,bm00b,um00,bcn00,zws99,as00b,btv02}. 

Currently, there is a tremendous amount of activity 
going on in trying to determine the equation of state $w_X(z)$ and 
other cosmological parameters from 
observations of high redshift Type Ia supernova 
\cite{gjc++98,astier00,srss00,wg01,wa01,gaagp01,wl01,leibundgut01,trentham01,pnr01,cc02,klsw02,mhh02,mbms02,wa02,ge02,rowan-robinson02,lj03,pc03,visser04,cmm04,alcaniz04,wm04,nbs03,kaa++03,zf03,dja04,gcd04,gc04,gd04,gong04,bertolami04,wt04,ccrl04,mcinnes04,sc04,bsss04,la04,np04,ass04,zfh04,ap04}. 
While there has been a considerable activity in this field, 
one should keep in mind that there are several theoretical 
degeneracies in the Friedmann model, which can 
limit the determination of $w_X(z)$. To understand 
this, note that the only 
non-trivial metric function in a Friedmann universe is the 
Hubble parameter $H(z)$ (besides the curvature of the spatial 
part of the metric), which is related 
to the \emph{total} 
energy density in the universe. 
\emph{Hence, it is not  possible to determine the energy
densities of individual components of energy densities in the 
universe from any geometrical observation.} For example, 
if we assume a flat universe,
and further assume that 
the only energy densities present are those corresponding to 
the non-relativistic dust-like 
matter and dark energy, then we need to know $\Omega_m$ of 
the dust-like matter and 
$H(z)$ to a very high accuracy in order 
to get a handle on $\Omega_X$ or $w_X$ of the dark energy. This can be 
a fairly strong degeneracy for determining $w_X(z)$ from observations.

Recently, we discussed  
certain questions related to the 
determination of the nature of 
dark energy component from observations of high redshift supernova 
in \citeN{pc03} [hereafter Paper~I].
In the above work, we
reanalyzed the supernova data using very simple 
statistical tools in order to focus attention on 
some key issues. The analysis of the data were 
intentionally kept simple as 
we subscribe to the point of 
view that any result which cannot be revealed by a simple 
analysis of data, but arises
through a more complex statistical procedure, is inherently 
suspect and a conclusion as 
important as the existence of dark energy with 
negative pressure should pass such a test. The key results 
of our previous analysis were: 

$\bullet$ Even if the precise 
value of $\Omega_X$ or the equation of state $w_X(z)$ is known from 
observations, it is {\em not} possible to determine the nature 
(or, say, the Lagrangian) of 
the unknown dark energy source using only kinematical
and geometrical measurements. 
For example, 
if one assumes that the dark energy arises from a scalar field, then 
it is possible to come up with
scalar field Lagrangians of different forms leading to same $w_X(z)$.
As an explicit example, we considered 
two Lagrangians, one corresponding to quintessence
\cite{pr88,rp88,zws99} and 
the other corresponding to the tachyonic scalar fields
\cite{padmanabhan02,pc02,fks02,sw02,gibbons02,ft02,mukohyama02,feinstein02,bjp03}. 
These two fields are quite different in terms of their intrinsic
properties; however,
it is possible to make both the Lagrangians 
produce a given $w_X(a)$ by choosing the potential functions in 
the corresponding Lagrangians [for explicit examples and forms 
of potential functions, 
see \citeN{padmanabhan02}; Paper~I].

$\bullet$ Although 
the full data set of supernova observations 
strongly rule out models without dark energy, 
the high and low  redshift 
data sets, individually, admit non-accelerating models with 
zero dark energy. It is not surprising that the high redshift data 
is consistent with non-accelerating models as the universe is in its 
decelerating phase at those redshifts. 
On the other hand, though the acceleration
of the universe is a low redshift phenomenon, the non-accelerating 
models could not be ruled out using low redshift data {\it alone} because of 
large errors.
Given the small data set, any possible evolution 
in the absolute magnitude of the supernovae, if detected, 
might have allowed the data to be consistent with the non-accelerating models.

$\bullet$ We introduced two parameters, which 
can be obtained entirely from theory,  
to study the sensitivity of the luminosity distance on $w_X$.
Using these two parameters, we argued that although 
one can  determine the present value of $w_X$ accurately from 
the data, one cannot constrain the evolution of $w_X$. The situation 
is worse if we add the uncertainties in determining $\Omega_m$.

All the above conclusions were obtained by analysing only 55 
supernova data points from a very simple point of view.
In recent times, data points from various sets of observations have been
compiled taking into account the calibration errors and other uncertainties. 
This enables us to repeat our analysis for much larger data sets, and 
see how robust are the conclusions of Paper~I with respect to 
the choice of the data points. In this paper, we will compare 
three such data sets, which differ in their selection criteria 
for data points and redshift range covered.

The structure of the paper is as follows: In the next section, we 
describe the three data sets used in this paper, 
and then analyse them for models 
with non-relativistic dust-like matter and 
cosmological constant. Some key points 
regarding the importance of low and high redshift data are discussed. 
In Section 3, we briefly discuss the constraints on the 
dark energy equation of state and its evolution. The results are 
summarized in Section 4. Finally, the effect of our 
extinction-based selection criterion on the determination of cosmological 
parameters is discussed in the Appendix.

\section{Recent supernova data and their analysis}
\label{sndata}

We begin with a brief outline of the method of our analysis
of the supernova data.
The observations essentially 
measure the apparent magnitude $m$ of a supernova at peak brightness
which, after correcting for galactic extinction and possible 
K-correction, is related to the 
luminosity distance $d_L$ of the supernova through
\be
m(z) = {\cal M} + 5 \log_{10} Q(z),
\label{mq}
\e
where 
\be
Q(z) \equiv \f{H_0}{c} d_L(z)
\e
and
\be
{\cal M} = M + 5 \log_{10} \left(\f{c/H_0}{1~\mbox{Mpc}}\right) + 25 
= M - 5 \log_{10} h + 42.38.
\e
The parameter $M$ is the absolute magnitude of the supernovae after correcting
for supernova light curve width - luminosity 
correlation \cite{rpk96,pgg++97,pls++99}.
After applying the above correction, $M$, and hence ${\cal M}$, is 
believed to be constant 
for all supernovae of Type Ia.

For our analysis, we 
consider three sets of data available in the literature at present. 
For completeness, we describe the data sets in detail:

(i) TONRY: In this data set we start with the 230 data points listed in 
\citeN{tsb++03} alongwith the 23 points 
from \citeN{btb++04}. 
These data points are compiled and calibrated from a wide range of different 
observations. 
For obtaining the best-fit cosmological model from the data, one 
should keep in mind that the very low-redshift points might be affected 
by peculiar motions, thus making the measurement of 
the cosmological redshift uncertain; hence we consider only those 
points which have $z > 0.01$. Further, since one is not sure about 
the host galaxy extinction $A_V$, we do {\it not} consider 
points which have $A_V > 0.5$.  
The effect of this selection criterion based on the extinction, is discussed 
in the appendix.
Thus 
for our final 
analysis, we are left with only 194 points 
[identical to what is used in \citeN{btb++04}], which 
is more than thrice compared to what was used in Paper~I. 

The supernova data points in \citeN{tsb++03} and \citeN{btb++04}
are listed in terms of the luminosity distance
\be
\mu_1(z) \equiv m(z) - {\cal M}_{\rm obs}(z) = 5 \log_{10} Q_{\rm obs}(z),
\label{mu1}
\e
alongwith 
the corresponding errors $\sigma_{\mu_1}(z)$. Note 
that the quantity $\mu_1(z)$ is obtained from observations
by assuming some value of ${\cal M}$. This assumed value
of ${\cal M}$ 
[denoted by ${\cal M}_{\rm obs}$ in equation (\ref{mu1})]
does {\it not} necessarily represent the ``true'' ${\cal M}$, and hence
one has to keep it as a free parameter while fitting the data.

Any model of cosmology will predict the 
theoretical value $Q_{\rm th}(z; c_{\alpha})$ with some undetermined parameters
$c_{\alpha}$
(which may be, for example, $\Omega_m, \Omega_{\Lambda}$). The best-fit model 
is obtained by minimizing the quantity
\be
\chi_1^2 = \sum_{i=1}^M \left[
\f{\mu_1(z_i) 
- {\cal M}_1 
- 5 \log_{10}Q_{\rm th}(z_i; c_{\alpha})}{\sigma_{\mu_1}(z_i)}
\right]^2
\label{chisq}
\e
where 
\be
{\cal M}_1 = {\cal M} - {\cal M}_{\rm obs}
\e
is a free parameter representing the difference between the 
actual ${\cal M}$ and its assumed value ${\cal M}_{\rm obs}$ 
in the data. To take into account the uncertainties arising because of 
peculiar motions at low redshifts, we add an
uncertainty of $\Delta v = 500$ km s$^{-1}$ to the distance error
\cite{tsb++03}, i.e.,
\be
\sigma_{\mu_1}^2(z) \to 
\sigma_{\mu_1}^2(z) + \left(\f{5}{\ln 10} 10^{-0.2 \mu_1}
\f{\Delta v}{c}\right)^2
\e
Note that this correction is most effective at low redshifts (i.e., 
for small values of $\mu_1$). The minimization of 
(\ref{chisq}) is done with respect to the parameter 
${\cal M}_1$ and the cosmological parameters 
$c_{\alpha}$.

(ii) RIESS(w/o HST): Recently, \citeN{rst++04} 
have compiled a set of supernova data 
points from various sources 
with reduced calibration errors arising from systematics. In 
particular, they have discarded various points 
from the TONRY data set where the classification of the supernova
was not certain or the 
photometry was incomplete -- it is claimed that 
this has increased
the reliability of the sample. The most reliable set of data, 
named as `gold', contain 
142 points from previously published data, plus 14 points 
discovered recently using HST \cite{rst++04}. 
Our second data set consists of 142 points from the above `gold' sample of 
\cite{rst++04}, which does {\it not} include 
the latest HST data [hence the name RIESS(w/o HST)]. 
Essentially, this
data set is similar to the TONRY data set in terms of the covered 
redshift range, but is supposed to be more ``reliable'' in terms 
of calibration and other uncertainties.

We would like to mention here that the data points in \cite{rst++04} are 
given in terms of the distance modulus
\be
\mu_2(z) \equiv m(z) - M_{\rm obs}(z),
\e
which differs from the previously defined quantity 
$\mu_1(z)$ in equation (\ref{mu1}) 
by a constant factor. Consequently, the $\chi^2$ is 
calculated from
\be
\chi_2^2 = \sum_{i=1}^M \left[
\f{\mu_2(z_i) 
- {\cal M}_2 
- 5 \log_{10}Q_{\rm th}(z_i; c_{\alpha})}{\sigma_{\mu_2}(z_i)}
\right]^2
\label{chisq2}
\e
where 
\be
{\cal M}_2 = {\cal M} - M_{\rm obs}
\e
Note that the errors $\sigma_{\mu_2}(z_i)$ quoted in \citeN{rst++04} 
already take into account the effects of peculiar motions.

(iii) RIESS: Our third data set consists of all the 156 points in the 
`gold' sample of \cite{rst++04}, which includes the latest points 
observed by 
HST. The main difference of this set from the previous two is that 
this covers the previously 
unpopulated redshift range $1 < z < 1.6$.

\begin{figure}
\begin{center}
\rotatebox{270}{\resizebox{0.45\textwidth}{!}{\includegraphics{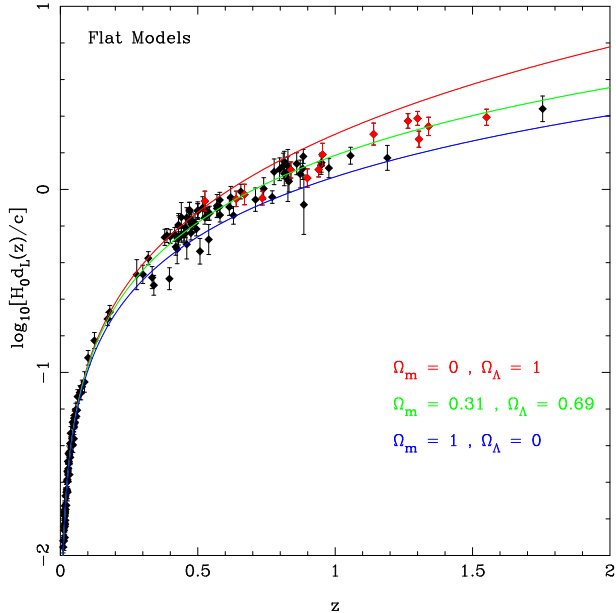}}}
\caption{Comparison between various flat models and 
the observational data. 
The observational data points, shown with error-bars, are obtained 
from the `gold' sample of Riess et al. (2004). The most recent
points, obtained from HST, are shown in red.
}
\label{dataflat}
\end{center}
\end{figure}

Before starting our analysis, we would like to caution the reader about 
two very important points. First, the errors $\sigma_m(z)$ used above do
not contain uncertainties because of systematics. Any rigorous 
statistical analysis of the supernova data for determining the 
cosmological parameters must take into account the systematic errors. 
The errors might arise because of calibration uncertainties, 
K-correction, Malmquist bias, 
gravitational lensing or the evolutionary effects
\cite{gma++02,gman02,l+h03,ps03,cmm04,hkkb04,linder04,wang04,klmm04}. 
Including 
such errors into the analysis requires much involved analysis. Once 
these systematic errors are included, the errors on the cosmological 
parameter estimations might be higher than what will be reported 
in this paper. In this respect, 
we would also like to add that the data sets RIESS and 
RIESS(w/o HST) are supposed 
to reduce some of the systematic and calibration uncertainties in data.

Second, our simple frequentist analysis 
holds good only when the errors $\sigma_m(z)$ are gaussian and 
uncorrelated. While considerable amount of analysis
exist in the literature working with these approximations, 
there are various systematics because of which such approximations 
do not hold true.
For example, the uncertainties in calibrating the data would surely
introduce correlations in the errors \cite{klmm04}. Similarly, uncertainties 
in the host galaxy extinction would introduce non-gaussian asymmetric 
errors. Neglecting such effects might result in lower errors on the 
estimated values of the cosmological parameters. 
Note that the main thrust of our analysis is to study some of the theoretical 
degeneracies inherent in any geometrical observations, in particular the 
supernova data, which are {\it not} adequately stressed elsewhere. 
Of course, this study can be complemented by other analyses 
which actually deal with quality and reliability of data, 
validity of error estimates, hidden correlations,
nature of statistical analysis etc.
All of these are important, but in order to make some key points 
we have attempted to restrict the domain of our exploration.
Keeping this in mind, we believe that the simple 
(non-rigorous) $\chi^2$ 
analysis should be adequate.

\begin{figure*}
\begin{center}
\rotatebox{270}{\resizebox{0.45\textwidth}{!}{\includegraphics{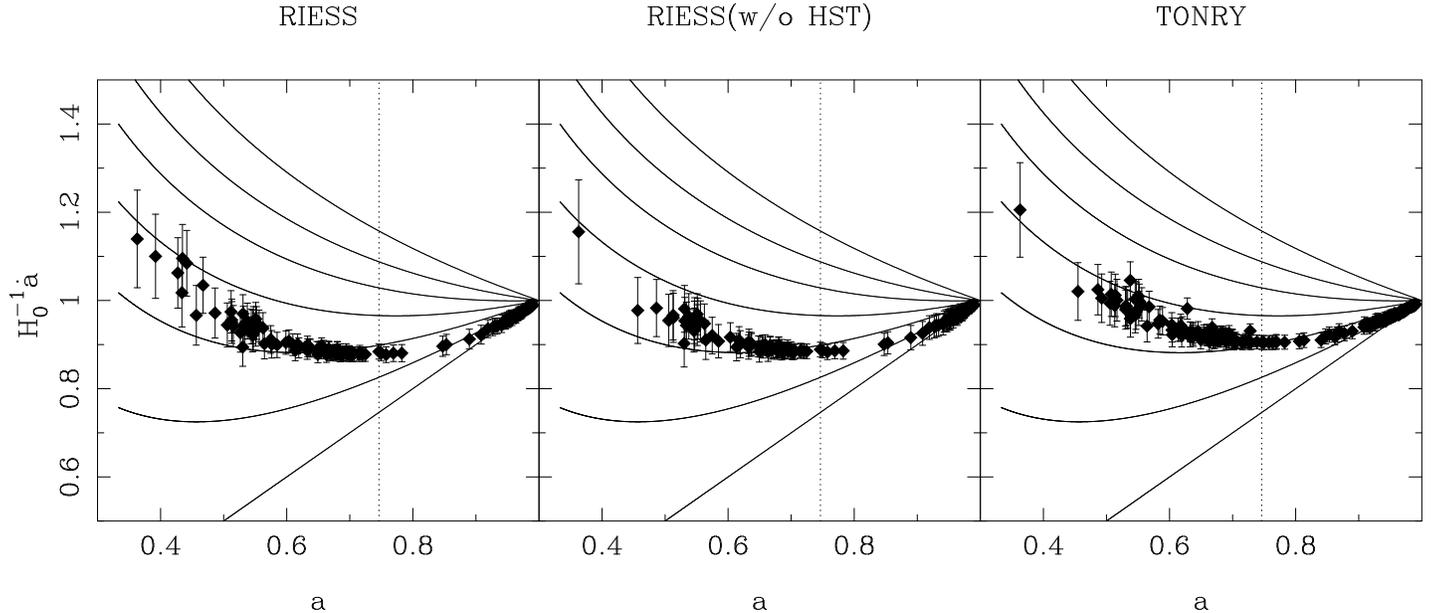}}}
\caption{The observed supernova data points in the $\dot{a} - a$ plane for 
flat models. The error bars for the data points
are correlated (see text for detailed description). The solid
curves, from bottom to top,  
are for flat cosmological models with 
$\Omega_m = 0.00, 0.16, 0.32, 0.48, 0.64, 0.80, 1.00$ respectively.
The left, middle and right panels show data points for the data sets 
RIESS, RIESS(w/o HST) and TONRY respectively. The vertical dashed line 
shows the redshift $z = 0.34$.}
\label{adotomegam}
\end{center}
\end{figure*}

\begin{figure*}
\begin{center}
\rotatebox{270}{\resizebox{0.8\textwidth}{!}{\includegraphics{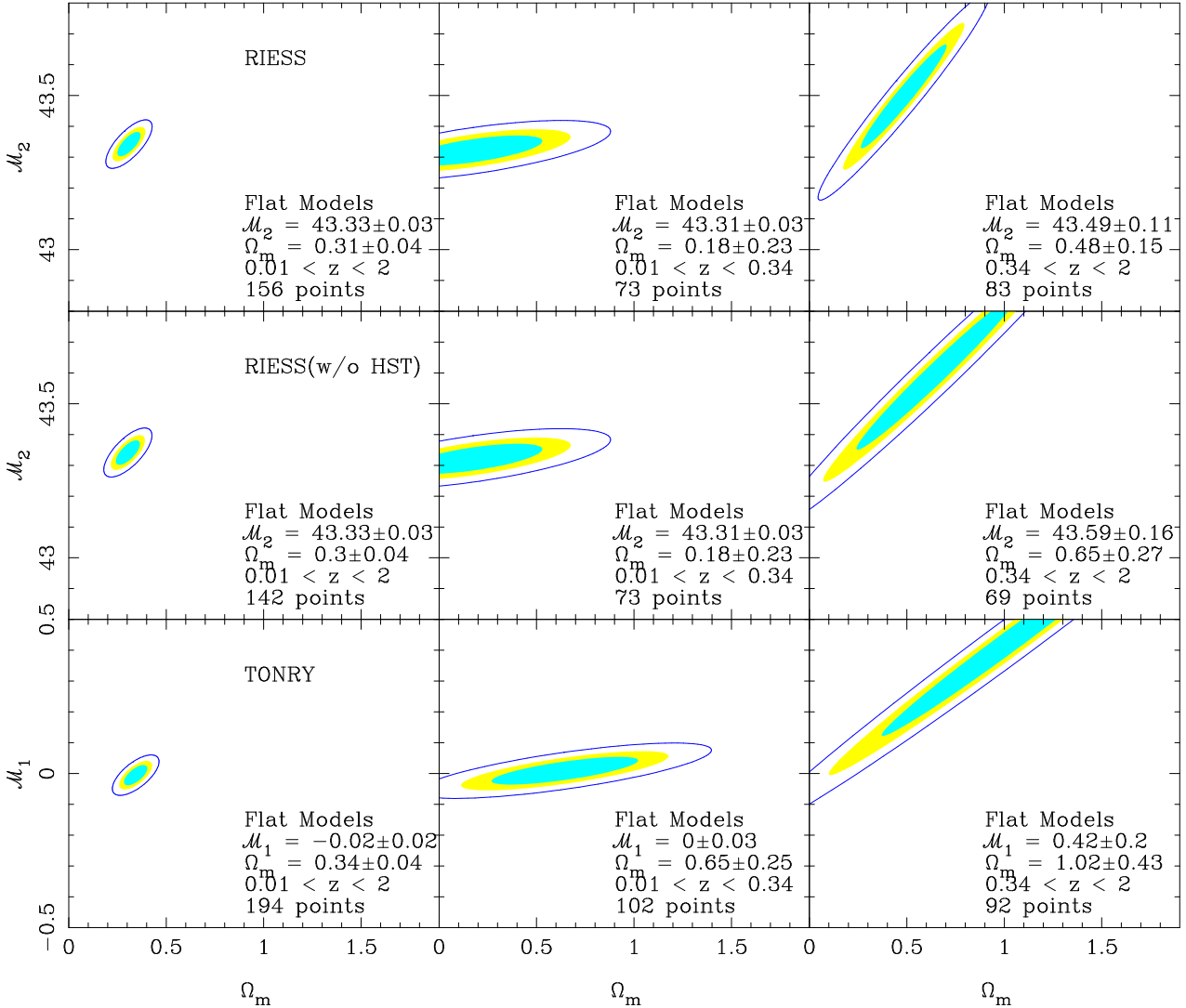}}}
\caption{Confidence region 
ellipses in the $\Omega_m - {\cal M}_{1,2}$ plane for flat models with 
non-relativistic matter and a cosmological constant. The 
ellipses corresponding to the 
68, 90 and 99 per cent confidence regions are shown.
The top, middle and bottom rows show data points for the data sets 
RIESS, RIESS(w/o HST) and TONRY respectively.
In the left panels, all the data points in the data set are used. 
In the middle panel, data 
points with $z < 0.34$ are used, while in the right panel, we have used  
data points with $z > 0.34$. We have indicated the best-fit values of 
$\Omega_m$ and ${\cal M}_{1,2}$ (with 1$\sigma$ errors).
}
\label{fitflat}
\end{center}
\end{figure*}
Let us start our analysis with the flat models where 
$\Omega_m + \Omega_{\Lambda} = 1$, 
which are currently favoured strongly 
by CMBR data (for recent WMAP results, see \citeNP{svp++03}). 
Our simple analysis for the most recent RIESS data set, with two free
parameters ($\Omega_m, {\cal M}_2$), gives 
a best-fit value of $\Omega_m$ (after marginalizing 
over ${\cal M}_2$) to be $0.31 \pm 0.04$
(all the errors quoted in this paper are 1$\sigma$). 
This matches with the value $\Omega_m = 0.29^{+0.05}_{-0.03}$
obtained by \citeN{rst++04}.
In comparison, the best-fit $\Omega_m$ for flat models 
was found to be $0.31 \pm 0.08$ in Paper~I -- thus there is  
a clear improvement in the errors because of increase in the 
number of data points although the best-fit value does not change.
The comparison between three flat models and 
the observational data from the RIESS data set 
is shown in in Figure \ref{dataflat}.

To see the accelerating phase of the universe more clearly, 
let us display 
the data as the phase portrait of the universe in the $\dot{a} - a$ plane. 
Though the procedure for 
doing this is described in Paper~I 
(see also \citeNP{dd03}), we 
would like to discuss some aspects of the procedure in 
detail to emphasize a different approach we have used 
here in estimating the errors. 

Each of the three sets of observational data used in this paper can be fitted 
by the function of simple form
\be
m_{\rm fit}(z) 
= a_1 + 5 \log_{10} \left[\f{z (1 + a_2 z)}{1 + a_3 z}\right],
\label{fitfunc}
\e
with $a_1, a_2, a_3$ being obtained by minimizing the $\chi^2$.
We can then represent the luminosity distance obtained 
from the data by the function
\be
Q_{\rm fit}(z) = 10^{0.2 [m_{\rm fit}(z) - {\cal M}]}
\e
Note that one needs to fix the value of ${\cal M}$ to 
obtain the function $Q_{\rm fit}(z)$. It is obvious, from the 
form of the fitting function (\ref{fitfunc}) at low redshifts, 
that the parameter $a_1$ actually measures the quantity ${\cal M}$. 
It is then straightforward to obtain
\be
Q_{\rm fit}(z) = \f{z (1 + a_2 z)}{1 + a_3 z}
\e
For flat models, it the Hubble parameter
is related to $Q(z)$ by a simple relation -- in this work 
we are interested in a related quantity
\be
H_0^{-1} \dot{a}(z) = \left[(1+z) \f{\de}{\de z} \left\{\f{Q(z)}{1+z}\right\}
\right]^{-1}
\e
which will enable us to plot the data points in the
$\dot{a} - a$ plane. Using the form of the fitting function, we can 
obtain the ``fitted'' $\dot{a}$ as:
\be
H_0^{-1} \dot{a}_{\rm fit}(z) = \f{(1 + a_3 z)^2 ~ (1 + z)}
{1 + 2 a_2 z + (a_2 - a_3 + a_2 a_3) z^2}
\label{adotfit}
\e

To plot the individual supernova data points in the $\dot{a} - a$ plane, we
first write $H_0^{-1} \dot{a}_{\rm fit}$ as a function of 
$m_{\rm fit}$ [which is trivially done by eliminating 
$z$ from equations (\ref{fitfunc}) and (\ref{adotfit})]. We then 
assume that the same relation can be applied to obtain the 
$\dot{a}$ corresponding to a particular measurement of $m$. 
Note that the relation between $\dot{a}$ and $m$ will involve 
the fitting parameters $a_1, a_2, a_3$, and hence is dependent 
on the fitting function.

The determination of the corresponding 
error-bars is a non-trivial exercise. In this paper, we obtain 
the error-bars using a Monte-Carlo realization technique, along the following lines:
Given the observed values of $m(z)$ and $\sigma_m(z)$, 
we generate random realizations of the data
set. Basically we randomly vary the magnitude of each supernova from a
gaussian distribution with dispersion $\sigma_m$ -- each such set corresponds
to one realization of the data set.
Next, we fit each of the realization of the 
data sets with the fitting function 
(\ref{fitfunc}), and obtain
the set of three parameters $a_1,a_2,a_3$. 
Given the set of parameters $a_1,a_2,a_3$, 
we can obtain $\dot{a}$ for each $a$ (or equivalently, $z$). In
this way we end up with different values of $\dot{a}$ for each supernova, each
corresponding to one realization.
Finally, we plot the distribution of $\dot{a}$'s for each supernova, fit
it with a gaussian, and obtain the width of the gaussian. This width is a
possible candidate for the error in $\dot{a}$ for each
supernova.

The data points, with error-bars, in the $\dot{a} - a$ plane are shown in 
Figure \ref{adotomegam} for all the three data sets. 
The solid curves plotted in Figure \ref{adotomegam} correspond 
to theoretical flat models with different $\Omega_m$.
In order to do any serious statistics 
with Figure \ref{adotomegam}, one should 
keep in mind that the errors for the data points in the figure are
correlated.

It is obvious that the high redshift data {\it alone} cannot 
be used to establish the existence of a cosmological constant as 
the points having, say $a < 0.75$, more or less, resemble 
a decelerating universe. 
In particular, 
one can use the freedom in the value of ${\cal M}$ to shift the 
data points vertically, and make them consistent with the non-accelerating 
SCDM 
model ($\Omega_m = 1$, topmost curve).
On the other hand, the low redshift 
data points show a clear, visual, sign of an accelerating universe 
at low redshifts. 
But to convert this visual impression into quantitative statistics 
is not easy since --- as we said before ---
the errors at neighbouring points are correlated. We shall see later on, 
with correct statistical analysis,
that it is, in general, 
quite difficult to rule out non-accelerating models using 
low redshift data alone, particularly when the uncertainties in the data 
are large.

\begin{figure*}
\begin{center}
\rotatebox{270}{\resizebox{0.8\textwidth}{!}{\includegraphics{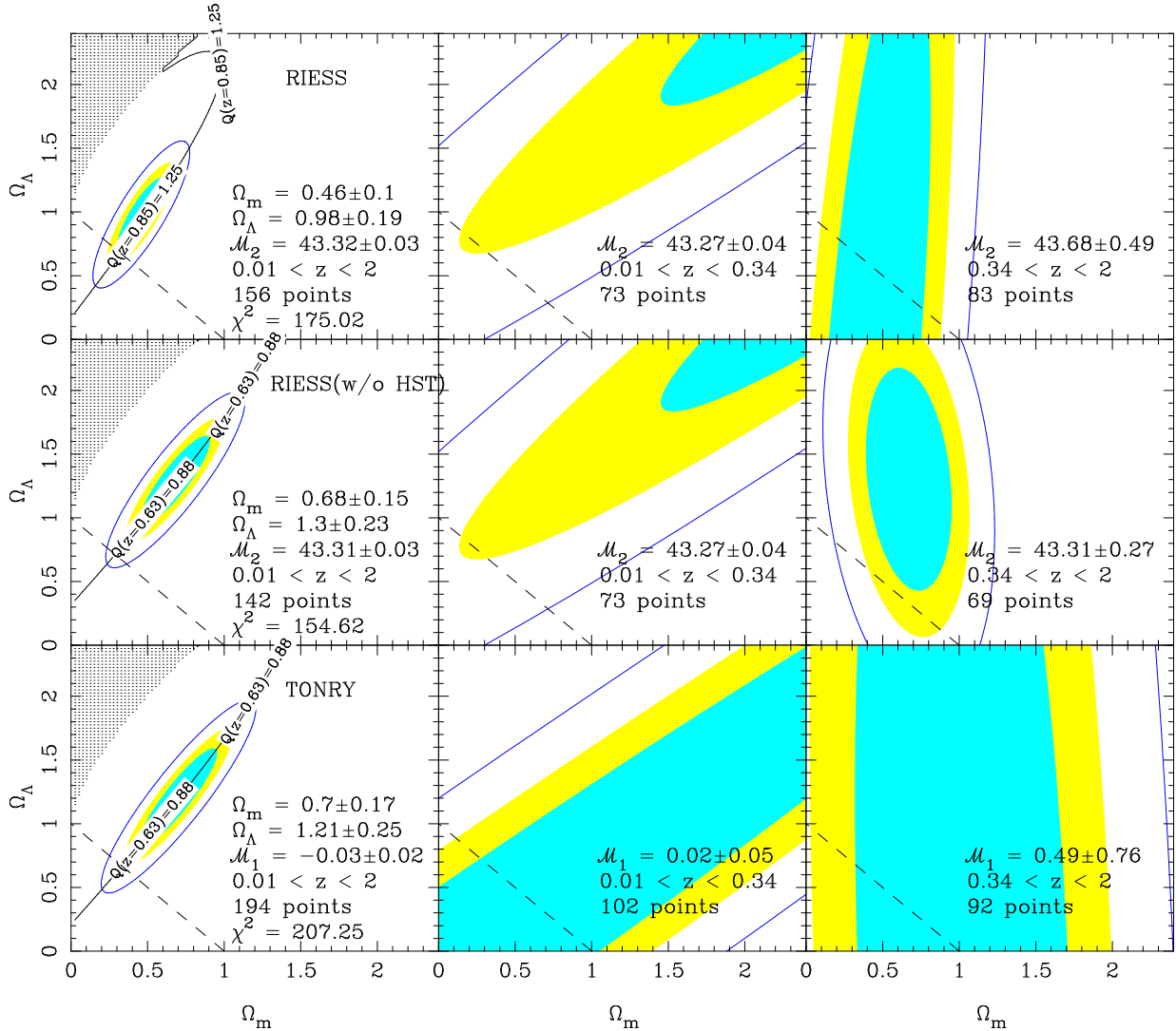}}}
\caption{Confidence region 
ellipses in the $\Omega_m - \Omega_{\Lambda}$ plane for models with 
non-relativistic matter and a cosmological constant. 
The ellipses corresponding to the 
68, 90 and 99 per cent confidence regions are shown.
The confidence regions are obtained after marginalizing 
over ${\cal M}_{1,2}$. The dashed line corresponds to the flat 
model $(\Omega_{~m} + \Omega_{\Lambda} = 1)$. 
The unbroken 
slanted line corresponds to the contour of 
constant luminosity distance, $Q(z) =$ constant. 
The top, middle and bottom rows show data points for the data sets 
RIESS, RIESS(w/o HST) and TONRY respectively.
In the left panels, all the data points in the data set are used. 
In the middle panel, data 
points with $z < 0.34$ are used, while in the right panel, we have used  
data points with $z > 0.34$. The values of the best-fit parameters, with 
1$\sigma$ errors are indicated in the respective panels.
}
\label{fitallz}
\end{center}
\end{figure*}

Let us now make the above conclusions more quantitative by 
studying the confidence 
ellipses in the $\Omega_m - {\cal M}_{1,2}$ 
plane, shown in Figure \ref{fitflat},
which should be compared with Figure 4 of Paper~I. 
For all the three rows, the left panels show the confidence regions using 
the full data sets. 
The confidence contours in the middle and right panels are obtained by 
repeating the best-fit analysis for the low redshift data set 
($z < 0.34$) and high redshift data set ($z > 0.34$), respectively.
\footnote{One might notice that, in Paper~I, 
we divided the high and low-redshift 
data points at $z = 0.25$, whereas in this paper we divide them 
at $z = 0.34$. The results of Paper~I remain unchanged irrespective 
of whether the points are divided at $z = 0.25$ or at $z = 0.34$; this 
is because there were very few points between these redshifts.}
The three rows are for the three data sets respectively, as indicated in 
the figure itself.

When the supernova data is divided into low and high redshift 
subsets, the points to be noted are:
(i) The best-fit value of ${\cal M}_{1,2}$ are substantially 
different for the two subsets (as indicated in the middle and right-hand 
panels of Figure \ref{fitflat}), irrespective of the 
data set used. The difference is most for the TONRY data set, 
comparatively less for the RIESS(w/o HST) data set and 
least for the RIESS data set.
(ii) Because of the difference in the value of 
${\cal M}_1$ for the TONRY data set, both the low and high 
redshift data subsets, when treated separately, are quite consistent with 
the SCDM model ($\Omega_m=1$). This 
indirectly stresses the importance of any evolutionary effects.
If, for example, supernova at $z\gtrsim 0.34$ and supernova at
$z\lesssim 0.34$ have different absolute luminosities because of
some unknown effect, or if there is any 
systematics involved in estimating the 
magnitudes of the supernova, then the entire TONRY 
data set can be made consistent
with the SCDM ($\Omega_m =1,\Omega_\Lambda =0$) model.
Comparing the best-fit values of ${\cal M}_1$ in 
the middle and right-hand panels in the lowest row 
of Figure \ref{fitflat}, one 
can see that a difference of about $0.5$ magnitude in the 
absolute luminosities of the low and high-redshift supernova 
is sufficient to make the entire TONRY 
data set consistent with the SCDM 
model. This agrees with the point made in Paper~I.
(iii) However, the situation is markedly 
different for the other two data sets [RIESS(w/o HST) and RIESS], 
which are supposed to be more reliable than the TONRY data set. 
It turns out that because of less systematic errors,  
it is possible to rule out the SCDM model using 
the low redshift data {\it alone} as long as 
the absolute luminosities of supernovae do not evolve 
within the redshift range $z < 0.34$. This is very important as it 
establishes the presence of {\it the accelerating phase of the universe 
at low redshifts irrespective of the evolutionary effects}. More 
reliable data sets at low redshifts will help in making this 
conclusion more robust.

Let us now consider the 
non-flat cosmologies where we have 
three free parameters, namely, $\Omega_m$, $\Omega_{\Lambda}$ and
${\cal M}_{1,2}$. 
The confidence region ellipses in the $\Omega_m$--$\Omega_{\Lambda}$
plane (after marginalizing over 
${\cal M}_{1,2}$) are shown in 
Figure \ref{fitallz}
for the three data sets.

The left panels, for all the three rows, give the confidence contours 
for the full data sets.
One can compare the equivalent panel (a) of Figure 5 in Paper~I 
with the left panels of Figure \ref{fitallz} and see that 
they are essentially similar. In the previous case the best-fit values 
for the full data set 
were given by $\Omega_m = 0.67 \pm 0.25, \Omega_{\Lambda} = 1.24 \pm 0.34$, 
which agree, within allowed errors, with the best-fit values 
(indicated in the figure itself) for all the three data sets.
The slanted shape of the probability 
ellipses in the left panels show that a particular linear combination of 
$\Omega_m$ and $\Omega_{\Lambda}$ is selected out by these observations 
(which
turns out to be $0.81 \Omega_m - 0.58 \Omega_{\Lambda}$ for the TONRY and 
RIESS(w/o HST) data sets, while it is $0.85 \Omega_m - 0.53 \Omega_{\Lambda}$ 
for the RIESS data set). 
This feature, of course, has nothing to do with supernova 
data and arises purely 
because the luminosity distance $Q$ depends strongly on  a 
particular linear combination of $\Omega_m$ and $\Omega_{\Lambda}$ 
\cite{gp95}. 
This point is illustrated by plotting the contour of 
constant luminosity distance, $Q(z) =$ constant in the left panels. 
The coincidence 
of this line (which roughly corresponds to $Q$ at a redshift 
in the middle of the data) with the probability ellipses 
indicates that it is the dependence of the luminosity 
distance on cosmological parameters 
which essentially determines the nature of this result.
This aspect was discussed in detail in Paper~I.

One disturbing aspect of all the three data sets (also 
noticed in the data sets right from the early days) is 
that the best-fit model favours a closed universe with 
$\Omega_{\rm tot} \equiv 
\Omega_m + \Omega_{\Lambda} > 1$. 
It is repeatedly argued  
that, due to the highly correlated nature of the probability contours
(indicated by the very elongated ellipses in the left panels 
of Figure \ref{fitallz}), the best-fit value 
does not mean much. While this is true, one can certainly ask what is the 
probability distribution for $\Omega_{\rm tot}$ if we marginalize over
everything else. 
Interestingly we get $\Omega_{\rm tot} = 1.91 \pm 0.41$ 
for the TONRY data set, 
$\Omega_{\rm tot} = 1.98 \pm 0.36$ for the RIESS(w/o HST) data set and 
$\Omega_{\rm tot} = 1.44 \pm 0.28$ for the RIESS data set.
Alternatively, one can also compute the probability 
${\cal P}(\Omega_{\rm tot} > 1)$ of obtaining $\Omega_{\rm tot} > 1$, which is 
found to be 
${\cal P}(\Omega_{\rm tot} > 1) = 0.97$ for the TONRY data set, 
${\cal P}(\Omega_{\rm tot} > 1) = 0.99$ for the RIESS(w/o HST) data set and 
${\cal P}(\Omega_{\rm tot} > 1) = 0.88$ for the RIESS data set.
Although there is a general consensus that the 
``concordance'' cosmological model 
has $\Omega_{\rm tot} = 1$, one should keep in mind that as 
far as supernova data {\it alone} is 
concerned, it is highly probable that $\Omega_{\rm tot} > 1$ 
--- in particular, 
the probability
is quite high ($\gtrsim 0.97$)  
when the recent HST data points are {\it not} included in the analysis.
The presence 
of 14 new HST points 
at redshifts around 1 to 1.6 makes sure that the probability 
of obtaining $\Omega_{\rm tot} > 1$ is somewhat lower ($< 0.9$).

\begin{figure*}
\begin{center}
\rotatebox{270}{\resizebox{0.3\textwidth}{!}{\includegraphics{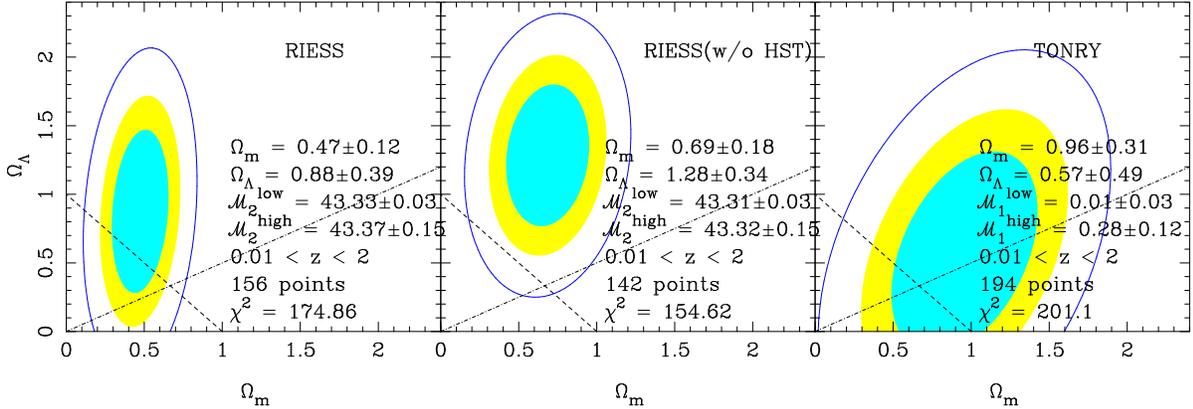}}}
\caption{Confidence region 
ellipses in the $\Omega_m - \Omega_{\Lambda}$ plane for models with 
non-relativistic matter and a cosmological constant, 
allowing for possibly different
${\cal M}_{1,2}$ for the different redshift subsamples. 
It is assumed that supernovae at $z < 0.34$ have ${\cal M}_{1,2}^{\rm low}$, 
while those at $z > 0.34$ have ${\cal M}_{1,2}^{\rm high}$.
The ellipses corresponding to the 
68, 90 and 99 per cent confidence regions are shown.
The confidence regions are obtained after marginalizing 
over ${\cal M}_{1,2}$. The dashed line corresponds to the flat 
model $(\Omega_{~m} + \Omega_{\Lambda} = 1)$. 
The dot-dashed line denotes the models having zero deceleration
at the present epoch 
(i.e., $q_0 = 0$), with the region below this line representing
the non-accelerating models. 
The left, middle and right panels show data points for the data sets 
RIESS, RIESS(w/o HST) and TONRY respectively.
The values of the best-fit parameters, with 
1$\sigma$ errors are indicated in the respective panels.
}
\label{fit2mu}
\end{center}
\end{figure*}

Finally, we comment on the interplay between high and low 
redshift data for non-flat models. Just as in the case of the flat models, 
we divide the full data set into low ($z < 0.34$) and high 
($z > 0.34$) redshift subsets, and repeat the best-fit analysis. 
The resulting confidence contours are shown 
in the middle and right panels of Figure \ref{fitallz}, which should 
be compared with panels (a) and (e) of Figure 7 in Paper~I.
One can see that 
it is not possible to rule out the SCDM model using 
only high redshift data points when there are large 
uncertainties in ${\cal M}_{1,2}$, which agrees with what we 
concluded in Paper~I.  
It is also clear that, like in Paper~I,  
the low redshift data for the TONRY data set cannot be used to discriminate 
between cosmological models effectively because of large errors on the data. 
However, the situation is quite different for the RIESS(w/o HST) and RIESS 
data sets. As we discussed before, the reduced uncertainties in these 
data sets have made it possible to rule out the SCDM model using 
low redshift data only. It is thus very important to have more data points 
at low redshifts (with less distance uncertainties) so as to 
conclude about the existence of accelerating phase of the universe, 
irrespective of evolutionary effects in absolute luminosities of supernovae. 

We also note, as we did for flat models, that the 
best-fit value of ${\cal M}_{1,2}$ are substantially 
different for the two subsets (as indicated in the middle and right-hand 
panels of Figure \ref{fitallz}) with  
the difference being most for the TONRY data set
and least for the RIESS data set.
We can thus take our analysis
one step further by fitting supernovae from all redshifts 
while allowing for possibly different
${\cal M}_{1,2}$ for the different redshift samples. To be precise, 
we assume that supernovae at lower redshifts 
$z < 0.34$ have ${\cal M}_{1,2}^{\rm low}$, 
while those at higher redshifts have ${\cal M}_{1,2}^{\rm high}$.
Given these, we can fit the data with four parameters 
and then marginalize over ${\cal M}_{1,2}^{\rm low}$ and 
${\cal M}_{1,2}^{\rm high}$. The resulting confidence regions 
in the $\Omega_m$--$\Omega_{\Lambda}$
plane are shown in 
Figure \ref{fit2mu}
for the three data sets. 

\begin{figure*}
\begin{center}
\rotatebox{270}{\resizebox{0.8\textwidth}{!}{\includegraphics{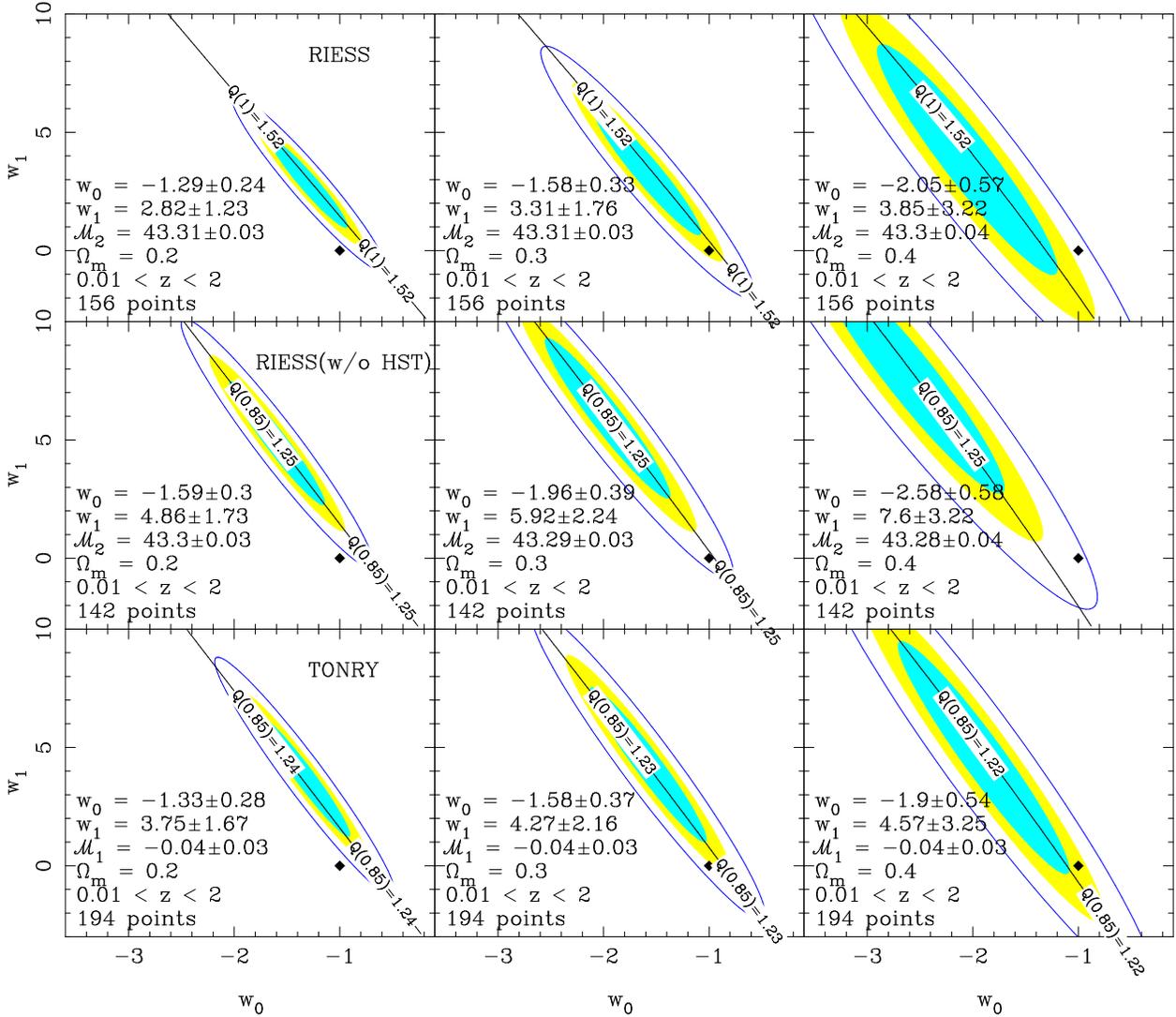}}}
\caption{Confidence region 
ellipses in the $w_{~0} - w_1$ plane for flat models with 
a fixed value of $\Omega_m$, as indicated 
in the frames. 
The confidence regions are obtained after marginalizing over 
${\cal M}_{1,2}$. The 
ellipses corresponding to the 
68, 90 and 99 per cent confidence regions are shown.
The square point denotes the equation of state 
for a universe with a non-evolving dark energy component (the 
cosmological constant). The unbroken 
slanted line corresponds to the contour of 
constant luminosity distance, $Q(z) =$ constant.
The top, middle and bottom rows show data points for the data sets 
RIESS, RIESS(w/o HST) and TONRY respectively.
The best-fit values of the fitted parameters $w_0$ and $w_1$ are indicated 
in the panels, alongwith the corresponding errors.
}
\label{fitw0w1}
\end{center}
\end{figure*}

As is clear from the figure, one has quite different values for 
${\cal M}_1^{\rm low}$ and ${\cal M}_1^{\rm high}$ for the TONRY
data set, while the difference is lower for the other two 
data sets. This probably indicates that the difference 
in the values of ${\cal M}_1$ for different subsets for the 
TONRY data set arises from systematic errors, which are claimed 
to be reduced for the other two data sets. One requires 
more work, possibly a rigorous study using Monte-Carlo simulations, 
to understand this in detail. 
One should also note that the 
data is consistent with the 
non-accelerating models 
at 68 and 99 percent confidence levels
for the TONRY and RIESS data sets respectively, while they 
are ruled out for the RIESS(w/o HST) data set.

Before ending this section, let us explain a subtle point in determining 
$\Omega_m$ and $\Omega_{\Lambda}$ from geometrical observations. 
As has been discussed in Paper I, the only 
non-trivial metric function in a Friedmann universe is the 
Hubble parameter $H(z)$ (besides the curvature of the spatial 
part of the metric), hence, it is not  possible to determine the energy
densities of individual components of energy densities in the 
universe from any geometrical observation.
However, the analysis in this section might give the wrong impression 
that we have actually been able to determine both 
$\Omega_m$ and $\Omega_{\Lambda}$ just from geometrical observations. 
The point to note that we have made a crucial {\it additional} 
assumption that the 
universe is dominated by non-relativistic matter and a cosmological 
constant, with known equations of state. Once this assumption 
about the equations of state is made, it allows us to determine 
the energy densities of the individual components. 
On the other hand, if, for example, 
we generalize the composition of the universe from a simple cosmological 
constant to a more general dark energy with unknown equation of state, it
will turn out that the constraints will become much weaker. We shall take
up this issue in the next section.

\section{Constraints on evolving dark energy}
\label{evoldarken}

As we have discussed in Paper~I, the supernova data can be used for 
constraining the equation of state of the dark energy. 
In this section, we shall examine the possibility of 
constraining $w_X(z)$ by comparing theoretical 
models with supernova observations. 

As done in Paper~I, we parametrize the function $w_X(z)$ in 
terms of two parameters $w_0$ and $w_1$:
\be
w_X(z) = w_0 - w_1 (a - 1) = w_0 + w_1 \f{z}{1+z},
\label{wxz}
\e
and constrain these parameters from observations. We shall 
confine our analyses to flat models in this section 
(keeping in mind that the supernova data favours a 
universe with $\Omega_{\rm tot} > 1$ when $w_0 = -1, w_1=0$).

If we assume $w_X$ does not evolve with time ($w_1 = 0$), then 
a simple best-fit analysis for RIESS data set shows that for a flat model with 
$\Omega_m = 0.31$ and ${\cal M}_2 = 43.34$ (the best-fit 
parameters for flat models, obtained in the previous section), 
the best-fit value of 
$w_0$ is $-1.03 \pm 0.07$ (which is nothing but the 
conventional cosmological constant). The data, as before in Paper~I, 
clearly rules out 
models with $w_0 > -1/3$ at a high 
confidence level, thereby supporting the existence of 
a dark energy component with negative pressure.  

One can extend the analysis to find the constraints in the 
$w_0 - w_1$ plane. 
As before, we limit our analysis to a flat universe. Ideally, one should
fit all the four parameters $\Omega_m, {\cal M}_{1,2}, w_0, w_1$, and 
then marginalize over $\Omega_m$ and ${\cal M}_{1,2}$ to obtain the 
constraints on $w_X$. 
However, if we put a uniform prior on $\Omega_m$ in the whole range, then
it turns out that it is impossible to get any sensible constraints 
on $w_0$ and $w_1$. Furthermore, we would like to present the results in 
such a manner so that one can see how the uncertainty in $\Omega_m$
affects the constraints on $w_X$.
Keeping this in mind, we fix the value of $\Omega_m$ 
to 0.2, 0.3 and 0.4 (which are typical range of values determined by 
other observations, like the LSS surveys, and are independent 
of the nature of the dark energy; \citeNP{pms++04,tsb++04,tbs++04}), and 
marginalize only over ${\cal M}_{1,2}$.

The 
confidence contours for the three data sets are shown in Figure \ref{fitw0w1}, 
which can be compared with Figure 8 of Paper~I.

The square point denotes the equation of state 
for a universe with a non-evolving dark energy component (the 
cosmological constant). 
The main points revealed by this figure are:
(i) The confidence contours are quite 
sensitive to the value of $\Omega_m$ used, thus 
confirming the fact (which was mentioned in Paper~I) that it 
is difficult to constrain $w_X$ with uncertainties 
in $\Omega_m$. For example, in the TONRY data set, 
we see that non-accelerating models 
with $w_0 < -1/3$ are ruled out with a high degree 
of confidence for low values of $\Omega_m$, while it is possible to
accommodate them for $\Omega_m \gtrsim 0.4$.
We have elaborated this point in Paper~I by studying the 
sensitivity of $Q(z)$ to $w_0$ and $w_1$ with varying $\Omega_m$.
(ii) The shape of the confidence 
contours clearly indicates that the data is not as sensitive to 
$w_1$ as compared to $w_0$. We stressed in Paper~I that 
this has nothing to do with the supernova data as such. Essentially, 
the supernova observations measure 
$Q(z)$ and it turns out that $Q(z)$ is 
comparatively 
insensitive to $w_1$. 
(iii) The best-fit values for all the 
three data sets strongly favour models with $w_0 < -1$, which
indicate the possibility of exotic forms of energy densities -- 
possibly scalar fields  
with negative kinetic energies [such models are explored, for example, in
\citeN{caldwell02}; \citeN{hm02}; \citeN{cht03}; \citeN{ckw03}; \citeN{mmot03}; \citeN{ssd03}; \citeN{johri04}; \citeN{stefancic04}; \citeN{st04}; \citeN{lh04}; \citeN{hl04}; \citeN{sck04}; \citeN{pz04}]. However, 
one should note that all the three data sets are still quite consistent 
with the standard cosmological constant 
($w_0=-1,w_1=0$) 
at 99 per cent confidence level for relatively 
higher values of $\Omega_m$. One still requires data sets of 
better qualities 
to settle this issue.
(iv) The inclusion of the new HST data points (RIESS data set) have resulted 
in drastic decrease in the best-fit 
value of $w_1$ (from 5.92 to 3.31 for $\Omega_m = 0.3$), 
implying less rapid variation 
of $w_X(z)$.

\section{Discussion}

We have reanalyzed the supernova data with the currently available data 
points and constrained various parameters related to 
general cosmological models and dark energy.
We would like to mention that our analysis ignores the effects of correlation 
and other systematics present in the data.
The main aim of the work has been to focus on some important theoretical issues which 
are not adequately stressed in the literature.
We have used three compiled 
and available data sets, which are called TONRY (194 points), 
RIESS(w/o HST) (142 points) and RIESS (156 points). The RIESS(w/o HST) is 
obtained from the TONRY data set by discarding points with 
large uncertainties and by reducing calibration errors, while the 
RIESS data set is obtained by adding the recent points from HST to the 
RIESS(w/o HST) set. The analysis 
is an extension to what was performed in Paper~I with a small subset 
of data points. 
In particular, we have critically compared the estimated values of 
cosmological parameters from the three data sets.  
While the errors on the parameter estimation
have come down significantly with all the data sets, we find that 
there some crucial differences between the data sets.
We summarize the key results once more:

$\bullet$ It has been well known that the supernova data 
rule out the flat and open matter-dominated models 
with a high degree of confidence 
\cite{rfc++98,pag++99,riess00}. However, for the TONRY and 
RIESS(w/o HST) data sets, we find that the data 
favours a model with $\Omega_{\rm tot} > 1$ 
(with probability $\gtrsim 0.97$) and 
is in mild 
disagreement with the ``concordance'' flat models
with cosmological constant.
This disagreement seem to be less (the 
probability of obtaining models with $\Omega_{\rm tot} > 1$ 
being $\approx 0.9$) 
for the RIESS data set,
which includes the 
new HST points in the redshift range $1 < z < 1.6$, 

$\bullet$ The supernova data on the whole
rules out non-accelerating models with very high confidence
level.
However, it is interesting 
to note that if we divide the TONRY data set 
into high and low redshift subsets, 
neither 
of the subsets are able to rule out the non-accelerating models. 
In particular, the low redshift data points are consistent with the 
non-accelerating models because of large errors on the data.
This keeps open the possibility that the evolutionary effects 
in the absolute luminosities of supernovae might make 
the entire data set consistent with SCDM model.  
The situation is quite different for the RIESS(w/o HST) and RIESS data sets, 
where points with large errors are discarded. 
The low redshift data 
alone seem to rule out the SCDM model with high degree of confidence. 
This means that unless the absolute luminosities of supernovae 
evolve rapidly with redshift, 
it might be difficult for the data set to be consistent
with the SCDM model. In other words, the RIESS(w/o HST) and RIESS data sets
establish the presence of the accelerating phase of the universe 
regardless of the evolutionary effects.

\begin{figure*}
\begin{center}
\rotatebox{270}{\resizebox{0.3\textwidth}{!}{\includegraphics{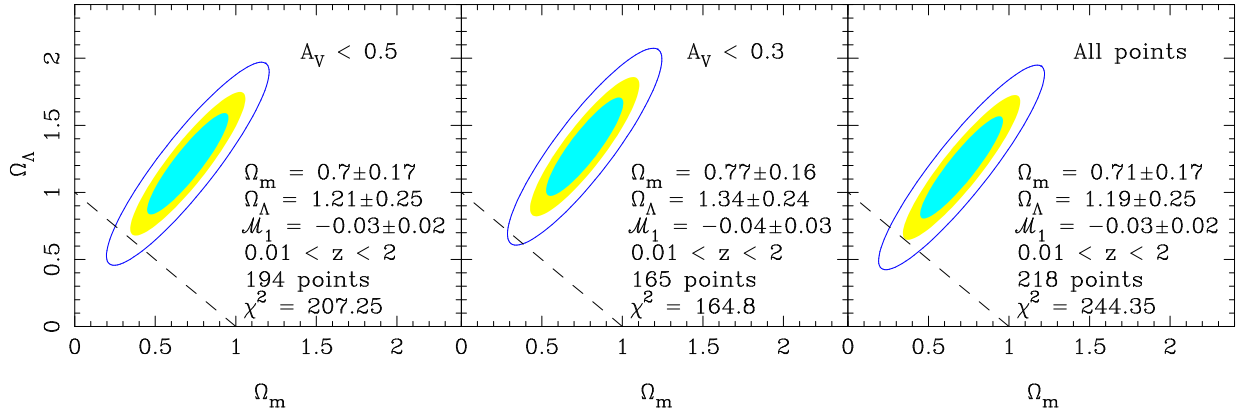}}}
\caption{Confidence region 
ellipses in the $\Omega_m - \Omega_{\Lambda}$ plane for models with 
non-relativistic matter and a cosmological constant for 
different selection criteria based on extinction for the TONRY data set. 
The ellipses corresponding to the 
68, 90 and 99 per cent confidence regions are shown.
The confidence regions are obtained after marginalizing 
over ${\cal M}_1$. The dashed line corresponds to the flat 
model $(\Omega_{~m} + \Omega_{\Lambda} = 1)$. 
The left panel shows results when only those points with 
$A_V < 0.5$ are included, the middle panel 
considers only points which have $A_V < 0.3$, while the 
right panel includes all the points irrespective of the value
of $A_V$.
The values of the best-fit parameters, with 
1$\sigma$ errors are indicated in the respective panels.
}
\label{fitav}
\end{center}
\end{figure*} 

$\bullet$
The key issue regarding dark energy is  
to determine the evolution of its equation of 
state, $w_X$. We find 
that although one can constrain the current 
value of $w_X$ quite well, it 
is comparatively difficult
to determine the 
evolution of $w_X$. The situation 
is further worsened when we take the uncertainties in $\Omega_m$ 
into account. 

$\bullet$ The supernova data {\it mildly} favours a dark energy equation of state 
with its present best-fit value less than -1 which will require more exotic forms of matter (possibly with 
negative kinetic energy). However, one should keep in mind that 
the 
data is still consistent with the 
standard cosmological constant at 99 per cent confidence level.

$\bullet$ The analysis of different subsamples of the supernova data set
is important in determining the effect of evolution. 
In this work, we have taken the simple approach 
of dividing the data roughly around the epoch where the 
universe might have transited from a decelerating to an 
accelerating phase, and checked whether the data can be made 
consistent with the non-accelerating models. In future, it 
would be interesting to divide the data based on the nature of 
supernova searches. For example, one can divide the data into 
three redshift splits: $z < 0.1$,
$0.2 < z < 0.8$ and  $z > 0.8$, 
which roughly correspond to supernovae discovered
in shallow searches, ground-based deep searches, and space-based deep
searches. It would be interesting to check the cosmological constraints 
with such a divide.

\section*{Acknowledgments}
We thank Alex Kim for extensive 
comments which significantly improved the paper.

\section*{Appendix: Effect of including supernovae with high extinction}

Since there is considerable uncertainty 
in determining the host extinction and reddening, we have considered 
only those supernova which have extinction $A_V < 0.5$ for the TONRY
data set. 
It would be interesting to see how this selection criterion affects 
our determination of cosmological parameters. In particular, one should
keep in mind that the high-redshift supernovae 
observed from the ground could have large
uncertainty in their color and hence statistically will often have
measured $A_V > 0.5$ even if they have no extinction.

To check how this affects the cosmological parameters, we concentrate 
on the cosmological models with non-relativistic matter and 
a cosmological constant, and find the constraints 
in the $\Omega_m - \Omega_{\Lambda}$ plane. We consider three 
cases, namely, (i) the usual one where we exclude all the data points 
with $A_V > 0.5$, (ii) the one with a stricter selection criterion where 
we exclude points with $A_V > 0.3$ and finally (iii) we include all 
the points irrespective of the extinction. The results for the 
three cases are plotted in Figure \ref{fitav}.
It is clear from the figure that the exclusion of points 
based on their extinction have little effect on the 
determination of the cosmological parameters, at least for the 
TONRY data set. The cosmological parameters agree within 1$\sigma$ errors
for the three different selection criteria.

\bibliography{aamnemonic,astropap}
 
\bibliographystyle{aa}

\end{document}